\documentclass{iau} 
\usepackage{graphicx}
\usepackage{natbib}

\title[Distant galaxies and massive stars] 
{What can distant galaxies teach us about massive stars?}

\author[E.~R.~Stanway]   
{Elizabeth R. Stanway$^1$
 }

\affiliation{$^1$Physics Department, University of Warwick, Gibbet Hill Road, Coventry, CV4 7AL, UK \\ email: {\tt e.r.stanway@warwick.ac.uk}}

\pubyear{2017}
\volume{329}  
\jname{The Lives and Death Throws of Massive Stars}
\editors{J.J. Eldridge, editor.}
\begin{document}

\maketitle

\begin{abstract}
Observations of star-forming galaxies in the distant Universe ($z>2$) are starting to confirm the importance of massive stars in shaping galaxy emission and evolution. Inevitably, these distant stellar populations are unresolved, and the limited data available must be interpreted in the context of stellar population synthesis models. With the imminent launch of JWST and the prospect of spectral observations of galaxies within a gigayear of the Big Bang, the uncertainties in modelling of massive stars are becoming increasingly important to our interpretation of the high redshift Universe. In turn, these observations of distant stellar populations will provide ever stronger tests against which to gauge the success of, and flaws in, current massive star models.
\end{abstract}

\firstsection 
\section{Introduction}

As studies elsewhere in these proceedings have shown, as many as $\sim$70\% of massive stars are expected to interact with a binary companion during their lifetimes. These stars dominate the integrated light of young stellar populations ($<100$\,Myr) and the effects of binary interactions are typically more pronounced at significantly sub-Solar metallicities.  This presents an interesting opportunity: the low metallicity, highly star-forming galaxies we observe at high redshifts both require an understanding of massive stellar evolution for their interpretation, and represent a laboratory in which to test that understanding. This synergy has recently been recognised by a growing subset of both the extragalactic and massive stars communities, and will only become stronger as the advent of highly-multiplexed near-infrared spectroscopy from the ground is complemented by the eagerly-anticipated {\em James Webb Space Telescope} (JWST).

Here I review recent observational indications of the presence and influence of massive stars in the distant Universe, as well as discussing the  the role of stellar modelling in their interpretation and the possible insights this synergy enables.

\section{Observing the Distant Universe}

\begin{figure}\begin{center}
\includegraphics[width=12cm]{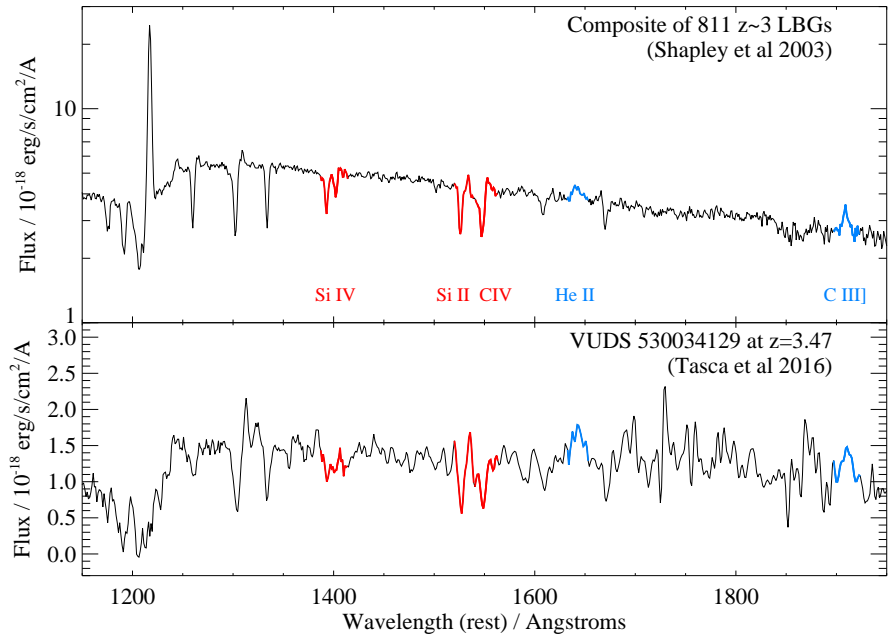}
\caption{Example spectra of high redshift sources, highlighting the features which confirm the presence of massive stars. Indicated lines are Si\,IV, Si\,II, C\,IV, He\,II and C\,III] - all lines which either include a direct component from massive stars or which are powered by the ionizing radiation of massive stellar populations. Early composites of Lyman break galaxies at $z\sim3$ \citep{2003ApJ...588...65S} have now been complemented by deep spectra of individual targets, the example given here being a bright $z=3.47$ source from the VUDS survey (Tasca et al 2016). \label{fig:spectra}}
\end{center}\end{figure}

Over the last twenty years, the number of spectroscopically-confirmed galaxies in the distant Universe (in this context, $z>2$) has grown exponentially, from a mere handful to tens of thousands of sources. The primary driver of this process is the `Lyman break technique' first applied on a large scale by \citet{1996ApJ...462L..17S}, and later extended to higher redshifts \citep[e.g.][]{2003MNRAS.342..439S,2011Natur.469..504B}. This method allows the selection of high redshift galaxy candidates by their distinctive photometric colours. These arise from a strong discontinuity imposed on their rest-frame ultraviolet spectrum due to absorption by neutral hydrogen in the intergalactic medium (IGM). It preferentially selects galaxies with high ultraviolet luminosity, and thus those with some component of on-going star formation. The redshift of these can be confirmed (for many cases) by follow-up spectroscopy of the rest-frame ultraviolet, redshifted into the observed optical. At very high redshift ($z\sim5-7$) spectroscopic characterisation is often restricted to detection of an isolated Lyman-$\alpha$ emission line ($\lambda_\mathrm{rest}=1216$\AA), with perhaps a second strong UV feature (e.g. He\,II or CIII]) to provide confirmation in rare cases \citep[e.g.][]{2015MNRAS.454.1393S}. Beyond $z\sim7$, the rising neutral hydrogen fraction in the IGM, combined with the shift of the Lyman break into the near-infrared, means that spectroscopic confirmation is very seldom possible, but the properties of galaxies may still be inferred from their photometry \citep[e.g.][]{2014MNRAS.443.2831C,2014ApJ...784...58S}.

\subsection{Rest-frame Ultraviolet Spectroscopy}
Rest frame ultraviolet spectroscopy can be interpreted through the construction of composites which yield the `typical' properties of galaxies in a population or subset thereof \citep[e.g.][]{2003ApJ...588...65S}. These have been complemented by  observations of individual sources, either particularly bright, or lensed targets, or simply using extremely deep spectroscopy \citep[e.g.][]{2016arXiv160201842T}.  While individual sources show variation, figure \ref{fig:spectra} illustrates certain features that are common in the high redshift galaxy population. Either directly (through features arising from the stellar spectra) or indirectly (through emission from the nebular gas of H\,II regions), these are diagnostic of the massive stellar population in these galaxies.

Clearly the strongest and most obvious of these features is the Lyman-$\alpha$ emission line. A full analysis of the emission and radiative transfer of this resonantly-scattered line is beyond the scope of this review. Nonetheless, both the equivalent width distribution of this line \citep[e.g.][]{2002ApJ...565L..71M,2007MNRAS.379.1589D,2017MNRAS.465.1543H}, and its typical offset in velocity from interstellar absorption features in the same source \citep[e.g.][]{2003ApJ...584...45A,2003ApJ...588...65S}, are indicative of a hard radiative field, powering galaxy-scale outflows. Given the young typical ages of galaxies (a few hundred Myrs), and the lack of evidence for significant AGN activity, these observations both imply the presence of sufficient massive stars to cause powerful radiative feedback and significantly affect the evolution of their circumgalactic environments.   

For populations in which we can move beyond Lyman-$\alpha$ (mostly at $z\lesssim5$, although with some exceptions), the rest-UV contains other diagnostics of massive stellar populations. As figure \ref{fig:spectra} shows, the prominant absorption features of C\,IV 1548,1550\AA, Si\,IV 1393,1402\AA\ and Si\,II 1526\AA\ are all seen in both composite and individual spectra. Similarly the He\,II 1640\AA\ and CIII] 1907,1909\AA\ emission features appear to be far more common in the distant galaxy population than in local star forming system. Each of these may have broad (stellar wind-driven) and narrow (nebular) components, but show strengths that are difficult to reproduce with conventional stellar populations \citep{2003ApJ...588...65S}. The emission lines in particular, are also indicative of a far harder ionizing spectrum than that seen in local sources. A few rare He\,II emitting star forming regions in the local Universe are interpreted as hosting massive, often Wolf-Rayet, stars \citep{2015ApJ...801L..28K}. Models incorporating more detailed analysis of massive stellar populations, either in terms of rotation or binary interaction, and exploring these effects at sub-Solar metallicities are proving both necessary for and successful in simultaneous fitting of these line strengths in the high redshift population \citep[e.g.][]{2012MNRAS.419..479E,2016ApJ...826..159S}.

\subsection{Rest-Frame Optical Spectroscopy}
\begin{figure}\begin{center}
\includegraphics[width=0.49\textwidth]{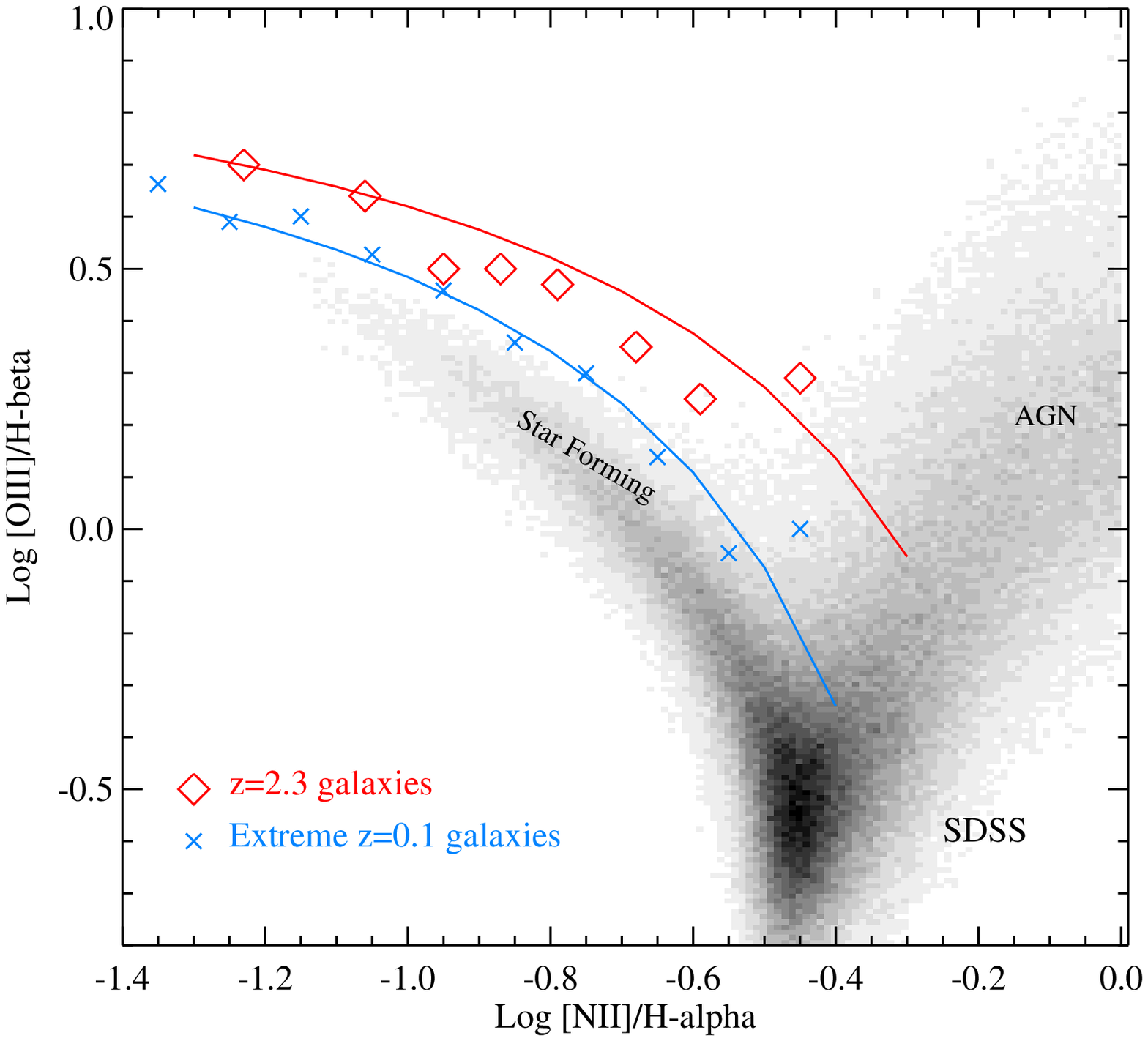}
\includegraphics[width=0.49\textwidth]{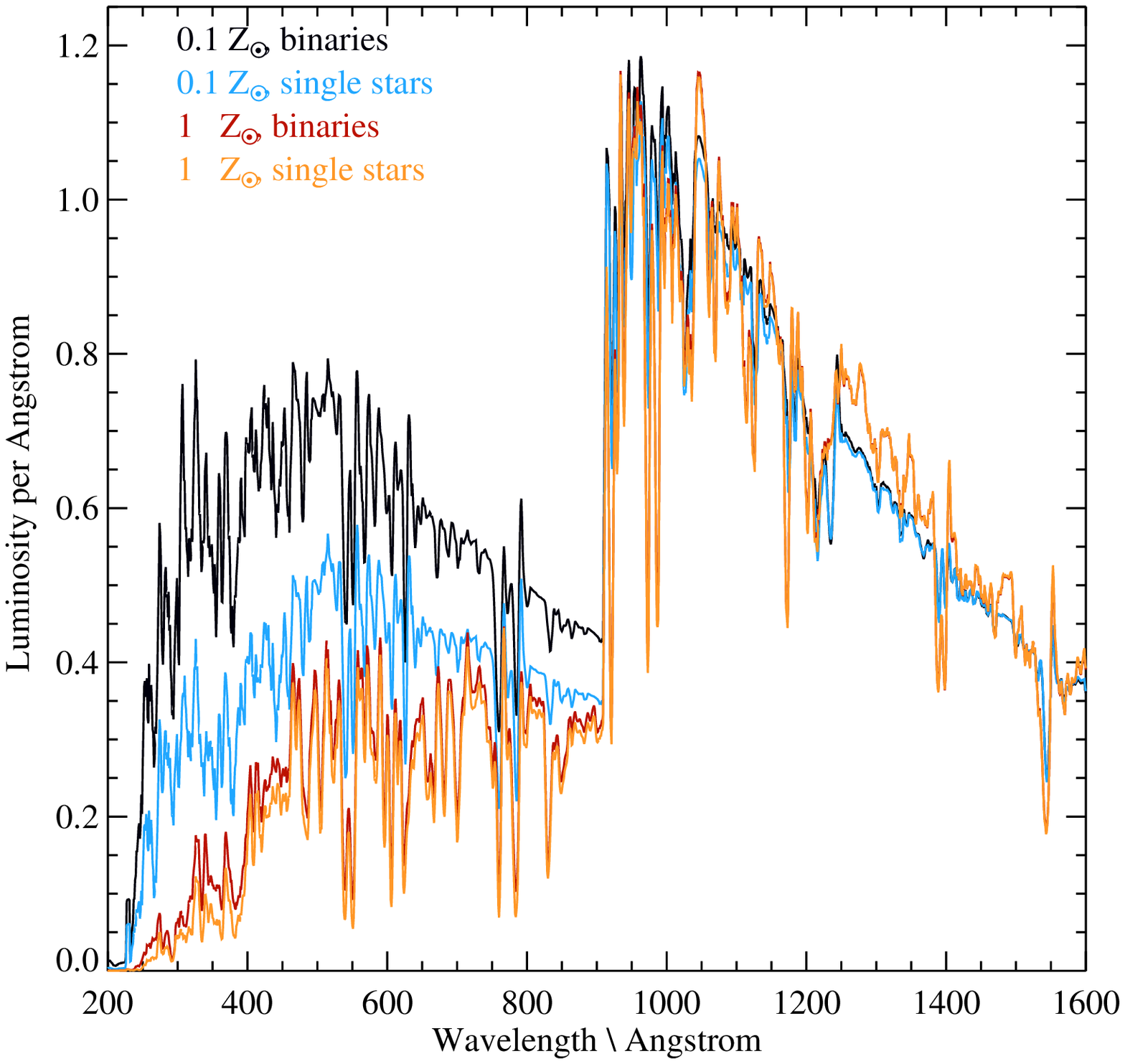}
\caption{The importance of the hardness of the ionizing spectrum. (Left) The often-used BPT diagram which measures the hardness of the ionizing spectrum both close to 1 Rydberg (the [NII]/H$\alpha$ ratio) and at 1-3\,Ryd ([OIII]/H$\beta$). Greyscale indicates the distribution of local star forming galaxies at near Solar metallicity from the SDSS. Points and lines indicate the properties of low metallicity, intensely star forming galaxies in the local Universe (blue crosses, Greis et al 2016) and at $z\sim2.3$ (red diamonds,  Strom et al 2016). (Right) The effect of metallicity and binary evolution effects on the hardness of the far-ultraviolet stellar continuum. Stellar population models have formed stars at a constant rate for 30\,Myr, and are taken from BPASS v2.0 (Stanway et al 2016).
  \label{fig:hardness}}
\end{center}\end{figure}

The recent advent of multi-object near-infrared spectrographs on 8-10m class telescopes has had a strong impact in this field.  The rest-frame optical spectra of $z\sim2-3$ Lyman break galaxies are now accessible in a reasonable integration time, allowing direct comparison with galaxy populations observed in the local Universe. In particular two programmes using the Keck/MOSFIRE instrument, KBSS \citep{2014ApJ...795..165S} and MOSDEF \citep{2015ApJ...801...88S}, have built large samples of several hundred nebular emission line spectra in this redshift range. As figure \ref{fig:hardness} (left) illustrates, one of the key findings of these programmes has been an offset in line ratios which probe the ionization conditions of nebular gas. The \citet[BPT,][]{1981PASP...93....5B} diagram is often used to distinguish between irradiation by star formation and that arising from active galactic nuclei. This diagram makes use of two pairs of lines, each close in wavelength to minimise the effects of dust and continuum subtraction uncertainties. The [N\,II]/H$\alpha$ ratio is primarily sensitive to the shape of the ionizing spectrum just above 1\,Rydberg, while the [O\,III]/H$\beta$ ratio probes the 1-3\,Rydberg range. The offset in the population medians for $z\sim2$ galaxies relative to the local $z<0.2$ Sloan Digital Sky Survey (SDSS) star forming galaxy population has been interpreted as evidence for a harder ionizing field than that of the near-Solar metallicity local population -- supporting a similar interpretation of the He\,II and C\,III] emission in the rest-frame ultraviolet \citep{2014ApJ...785..153M,2014MNRAS.444.3466S,2016ApJ...816...23S,2016ApJ...826..159S,2016arXiv160802587S}.

These high ionization environments are not unique to the highest redshifts. A number of $z\sim0-0.3$ galaxy populations have been identified as analoguous to those at high redshift, in terms of their star formation  and emission properties \citep[see][]{2016MNRAS.459.2591G}. As figure \ref{fig:hardness} demonstrates, these also appear distinct from the bulk of the low redshift galaxy locus in the BPT diagram, and so may provide more local (and hence accessible) environments in which to test our understanding of these stellar populations.

\subsection{Indirect Constraints}

At the highest redshifts, detailed spectroscopy is seldom possible. Nevertheless, there are strong indications that the dominant stellar population differs from that seen either at $z\sim2-4$ or in the local Universe. The rest-frame ultraviolet photometric colours of galaxies at $z>5$ indicate extremely blue spectral slopes; again, these are difficult to fit with most stellar population synthesis models unless a very young, very low metallicity population is invoked \citep[e.g.][]{2005MNRAS.359.1184S, 2012ApJ...756..164F, 2016MNRAS.455..659W}. As photometry extends to the rest-frame optical (shifted to the thermal infrared, and necessarily observed from space), spectral energy distribution modelling has suggested that many distant galaxies require extremely powerful nebular emission lines - consistent with the hard ultraviolet slope \citep[e.g.][]{2014ApJ...784...58S}.

A second indirect constraint is derived from the so-called Epoch of Reionization, a period at $z>7$ over which the IGM was ionized by ultraviolet photons escaping from star forming galaxies. Current constraints on the evolution of the ionized hydrogen fraction require either that the luminosity function of galaxies extends to very low stellar masses, that the escape fraction of photons is unreasonably high, or that the ionizing photon output of galaxies exceeds that currently inferred from their 1500\AA\ (rest) luminosities. All three factors are probably significant, but it appears likely that a hard ionizing spectrum, arising from evolution in the massive stellar population of typical galaxies, is a key requirement \citep[][]{2015ApJ...800...97T,2016MNRAS.456..485S,2016MNRAS.458L...6W,2016MNRAS.459.3614M}.

\section{The Key Role of Population Synthesis Models}\label{sec:bpass}

When observing distant galaxies, it is very rare that individual
stars, or even individual star forming regions can be resolved. At
$z=1$, the angular scale is $\sim$8\,kpc per arcsec and sub-kiloparsec
resolution is usually only possible from space, and for cases
distorted by strong gravitational lensing. As a result, the emission
that we observe is the integrated light of large stellar populations,
potentially of varying age and metallicity.  As the previous section
demonstrates, interpreting this necessarily requires comparison with
stellar population synthesis (SPS) models, which construct the
expected integrated spectrum from a stellar population with known
parameters (star formation history, initial mass function (IMF),
metallicity etc.) for comparison with the data. A number of SPS codes
are in common usage. These range from those based on purely empirical
observations of nearby stellar populations to others based on
theoretical stellar evolution and atmosphere models. Each contains
different formalisms for evolutionary stages and the handling of
complex populations, and as a result, each has different strengths and
weaknesses. Recent reviews of a number of SPS codes include those by
\citet{2013ARA&A..51..393C} and \citet{2016MNRAS.457.4296W}.

Here I will focus on results from a particular SPS model set - the Binary Population and Spectral Synthesis code \citep[BPASS, ][Eldridge et al in prep]{2009MNRAS.400.1019E,2012MNRAS.419..479E,2016MNRAS.456..485S}\footnote{see bpass.auckland.ac.nz}. This uses a library of theoretical stellar evolution models, and combines these with synthetic stellar atmospheres. Importantly, it also includes prescriptions for binary system evolutionary effects omitted in many other SPS codes, including mass loss and (a simple prescription for) rotation. At Solar metallicity, the output model spectra for continuously star forming populations are comparable to those of other codes for a Salpeter-like initial mass function, while at early ages, different IMFs and low metallicities the different model sets diverge. The effects of binary evolution tend to prolong the epoch over which very blue stars dominate the spectrum. The results presented here are from the v2.0 BPASS data release, supplemented by lower metallicity models which will be made available in BPASS v2.1, and for an IMF with a broken power law slope, with indices -1.35 at M$<0.5$\,M$_\odot$ and -2.35 for $0.5\,$M$_\odot<$M$<$M$_\mathrm{max}$.

\begin{figure}\begin{center}
\includegraphics[width=12cm]{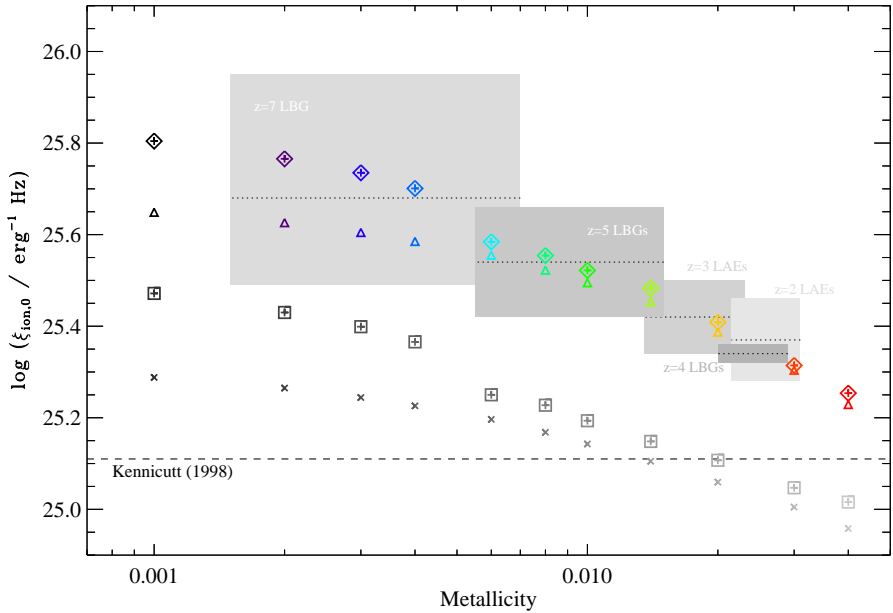}
\caption{The dependence of ionizing photon efficiency, $\xi_\mathrm{ion,0}$, on metallicity, initial mass function and single vs binary star evolution, assuming $f_{esc}=0$. At each metallicity we show the model values for binary (large, crossed symbols) and single star (smaller symbols) stellar populations, and for two IMFs (M$_\mathrm{max}$=300\,M$_\odot$, upper pair, and M$_\mathrm{max}$=100\,M$_\odot$, lower pair). Shaded regions indicate observational constraints ($\pm 1\sigma$) on $\xi_\mathrm{ion,0}$ from high redshift galaxy populations from Bouwens et al (2016, $z\sim4-5$ LBGs), Matthee et al (2017, $z=2$ LAEs, using the $\beta$ dust correction for consistency with Bouwens et al), Nakajima et al (2016, $z=3$ LAEs)  and Stark et al (2015, $z\sim7$ LBG). Metallicity is poorly constrained in these observations and the constraints are shown offset for clarity. Models give values for a 100\,Myr old, continuously star forming population using BPASS v2.0 (Stanway et al, 2016).
  \label{fig:ionizing}}
\end{center}\end{figure}

Such models can confront the very high ionizing fluxes inferred for galaxies in the distant Universe. In figure \ref{fig:ionizing} we illustrate the dramatic difference in ionizing flux output between different stellar populations and compare these to observational constraints derived from distant sources. These are usually presented using the parameter $\xi_\mathrm{ion} = \dot{N}_\mathrm{ion}/L_\mathrm{UV}$, i.e. the ratio between the ionizing photon production rate and the rest-frame ultraviolet continuum luminosity. While the latter can be directly observed in the distant Universe, the former must be inferred from indirect observations - primarily of the strong nebular emission lines generated from the galaxy in question, assuming that a fraction ($1-f_{esc}$) of the ionizing continuum is absorbed by the intertellar medium \citep[see][for discussion]{2015MNRAS.454.1393S,2016ApJ...831..176B}. While the metallicity of stellar populations in the distant Universe are still poorly constrained, the inferred ionizing fluxes are significantly higher than those predicted for a canonical, continuously forming stellar population at near-Solar metallicity \citep[the canonical value of][is indicated on the figure]{1998ARA&A..36..189K}. Instead, populations with a higher upper mass cut-off (300\,M$_\odot$ rather than the usually-assumed 100\,M$_\odot$) and those which account for binary effects (e.g. through evolutionary effects or rotation) are favoured. The steady increase in  $\xi_\mathrm{ion}$ measurements with redshift is more rapid than that expected from simple cosmic metallicity evolution for single star models, and may indicate that binary effects at low metallicity are becoming significant.

Not only the ionizing photon flux but also the hardness of its spectrum can provide insight into the massive stellar populations in distant galaxies.  In figure \ref{fig:hardness} (right) we illustrate the shape as well as the strength of the Lyman continuum emission region for stellar models with the same star formation history (constant star formation over a 30\,Myr interval) but with different metallicities and including or omitting the effect of stellar binaries. The shape of the ionizing spectrum is quite different, and will result in shifts in the ratios of line emission, since these probe different energy ranges. In particular, the presence of a hard (blue) ionizing spectrum in the 1-3\,Rydberg range will shift galaxies vertically in the ionization-sensitive BPT diagram (see also Xiao et al 2017, in these proceedings). If the observed shift in the populations of both distant galaxies and their local analogues are interpreted as a stellar population effect then models suggest that low metallicity models which incorporate binary evolution effects are required to reproduce them \cite[although note that models with enhanced N/H ratios may also be appropriate,][]{2015ApJ...801...88S}.

\section{Implications}\label{sec:implications}

There is clear evidence that the rest-frame ultraviolet spectra of galaxies, and the nebular emission powered by reprocessing of the same photons, have evolved over cosmic time. This is, of course, to be expected - the volume-averaged mean metallicity of the IGM and the mean stellar population age both drop towards higher redshifts simply because the Universe itself is younger. The observed metallicity of gas clouds in the IGM evolves slowly, with the matter density in C\,IV absorbers dropping by a dex between $z\sim2$ and $z\sim6$ \citep{2013MNRAS.435.1198D}. This is sufficient that we would expect to see significant changes in the character and influence of the massive star population over the same epoch. At the relatively low metallicities prevalent in the distant Universe, and the still lower ones inferred for the first few generations of star formation (i.e. during the Epoch of Reionization), we would expect the stars to be more massive, and to retain more of their mass due to the weakening of stellar winds. At the same time the influence of binary interactions will be important for these swollen massive stars. They are likely to exchange mass and angular momentum with binary companions, and so to rejuvenate ageing secondary stars. They may also spin up to velocities where rotational mixing influences the stellar evolution pathway. These changes are reflected in the far higher ionizing photon output seen in figure \ref{fig:ionizing} at low metallicities, particularly for populations with a slightly higher limiting stellar mass than previously assumed.

Figure \ref{fig:ionizing} also demonstrates why observers of distant galaxy populations are increasingly turning towards synthesis codes incorporating detailed stellar models to interpret their data. Constraints on the ionizing photon production in distant galaxies are suggesting that the effects of both IMF and multiplicity are becoming impossible to ignore in these sources. A corollary of this is that such galaxies provide a laboratory in which to test our models for massive stellar populations - they place the effects of these massive stars into a wider context. If a set of stellar population synthesis models cannot reproduce the observed properties of a distant source, we are left with three alternatives: either the observations are incorrect, or the wrong type of object is being modelled, or the models need to be reviewed and improved. In the local Universe it has rarely been necessary to answer such questions: the stellar population can frequently be resolved, or observed in sufficient detail to remove any ambiguity, while the models of near-Solar stellar populations are relatively mature and well constrained. The distant Universe, on the other hand, is now throwing up cases where searching questions can be asked of those modelling stellar populations, primary amongst them `do we understand how massive stars form, interact and evolve at low metallicity?'.

A case in point is that of the $z=6.6$ Lyman break galaxy CR7. This source was selected for its extremely strong Lyman-$\alpha$ line emission, but also shows strong He\,II 1640\AA\ emission, with constraints on the non-detection of O\,III 1665\AA\ and C\,III] 1909\AA\ lines \citep{2015ApJ...808..139S}. Together with its photometric colours, these characteristics have been proposed as indicative of a metal-free (Population III) stellar population, while alternative interpretations include a primordial Direct Collapse Black Hole \citep{2015MNRAS.453.2465P}. Either would be somewhat surprising, given the source's luminosity and redshift. As \citet{2016arXiv160900727B} discuss, stellar evolution models can go some way towards addressing this question, but are unable to resolve the issue completely and the interpretation of CR7 will remain ambiguous until further observations and improved modelling of all possible scenarios are undertaken.

While this is an isolated case, and the constraints placed on massive stellar models by galaxies (rather than the reverse) are still weak, JWST will revolutionise this field. The NIRSPEC instrument will enable deep spectroscopy of hundreds (if not thousands) of star-forming galaxies at $3<z<8$, from the rest-frame ultraviolet through to the rest-optical \citep[see e.g.][]{2016ASPC..507..305G}. As a result, it is likely to produce both more anomalous examples and much tighter constraints on the metal abundance and interstellar medium properties in galaxies in the distant Universe. Direct measurements of strong line ratios and probes of the stellar and interstellar absorption lines (as opposed to just strong line emission) should be possible on both individual and stacked sources. These will allow direct comparison with stellar population synthesis models, and test whether these recover the observed parameter spaces. Whether our understanding of the massive star population and its behaviour is sufficient to confront the wealth of observational data expected in the next few years remains to be seen. It will certainly be tested. Evidence to date is that both fields can learn from and be enriched by this synergy and it is my hope that they will continue to do so.

\end{document}